\documentclass[conference]{IEEEtran}
\IEEEoverridecommandlockouts
\usepackage{cite}
\usepackage{amsmath,amssymb,amsfonts}
\usepackage{algorithmic}
\usepackage{graphicx}
\usepackage{textcomp}
\usepackage{xcolor}
\usepackage{tabularx}
\usepackage{tikz}
\usepackage{float}
\usepackage{booktabs}
\usepackage{cuted}
\usepackage{amssymb}
\usepackage{pifont}
\usepackage{array}
\usepackage{algorithm}
\usepackage[most]{tcolorbox}
\usepackage{listings}
\definecolor{lightgray}{gray}{0.95}
\lstdefinestyle{prompt}{
    basicstyle=\ttfamily\footnotesize,
    breaklines=true,
    frame=single,
    numbers=none,
    backgroundcolor=\color{gray!8},
    keywordstyle=\bfseries,
    showstringspaces=false,
    columns=flexible
}

\newcommand{\cmark}{\textcolor{green!60!black}{\ding{51}}}
\newcommand{\xmark}{\textcolor{red}{\ding{55}}}

\definecolor{takeawaygray}{RGB}{236, 236, 236}

\newcounter{boxcounter}

\usetikzlibrary{positioning,fit,calc,arrows.meta}
\usetikzlibrary{positioning, arrows.meta, shapes.geometric, fit}
\def\BibTeX{{\rm B\kern-.05em{\sc i\kern-.025em b}\kern-.08em
    T\kern-.1667em\lower.7ex\hbox{E}\kern-.125emX}}
\pgfdeclarelayer{background}
\pgfsetlayers{background,main}
\begin{document}

\title{From Detection to Response: A Deep Learning and Retrieval-Augmented Generation Framework for Network Intrusion Mitigation
}

\author{
\IEEEauthorblockN{Md Navid Bin Islam}
\IEEEauthorblockA{
\textit{University of Northern British Columbia}\\
Prince George, Canada \\
navid.islam@unbc.ca}

\and

\IEEEauthorblockN{Sajal Saha, Senior Member (IEEE)}
\IEEEauthorblockA{
\textit{University of Northern British Columbia}\\
Prince George, Canada \\
sajal.saha@unbc.ca}
}
\maketitle

\begin{abstract}
Machine-learning-based Intrusion Detection Systems (IDS) have achieved impressive accuracy in classifying network attacks, yet they consistently fall short on the question that matters most to a security analyst: \emph{what should I do next?} This paper presents a unified, end-to-end framework that closes the gap between threat detection and actionable response. The system operates in two tightly coupled stages. First, an ensemble of three independently trained binary Deep Neural Networks (DNNs) \cite{fukushima1980neocognitron} classifies network traffic flows as Benign, Denial of Service (DoS) \cite{hussain2003framework}, or Distributed Denial of Service (DDoS) \cite{mirkovic2004taxonomy}, achieving 99.84\% accuracy on the CICIDS2018 \cite{Sharafaldinetal2018} dataset and 95.30\% on the UNSW-NB15 \cite{MoustafaSlay2015} dataset. Second, a Retrieval-Augmented Generation (RAG) \cite{lewis2020retrieval} pipeline constructs explanation-aware prompts from the top-5 anomalous features, retrieves the most semantically and lexically relevant guidance from a knowledge base derived from authorized sources and directs a locally deployed language model to synthesise structured, citation-grounded mitigation reports. The RAG-enhanced reports outperform vanilla LLM outputs across all automated evaluation metrics.
\end{abstract}


\section{Introduction}
Modern computer networks are experiencing a rapid increase in cyber threats, especially  Denial of Service (DOS) \cite{hussain2003framework} and Distributed Denial of Service (DDoS) \cite{mirkovic2004taxonomy} attacks. Recent industry reports show that more than 7.9 million DDoS incidents were recorded globally during the first half of 2024 alone \cite{john2024ddos}. As cyber-attacks continue to grow in scale and complexity, machine learning-based Intrusion Detection Systems (IDS) have become widely used because of their ability to detect attacks with very high accuracy. However, high detection accuracy alone is not sufficient in real-world cybersecurity operations.

In practice, detecting an attack is only the first step. Security analysts in a Security Operations Center (SOC) also need to understand the nature of the attack, identify the affected systems, and decide which mitigation actions should be taken immediately. A prediction label such as ``DDoS'' does not provide enough information to support rapid decision-making during a live incident. The delay between attack detection and effective response can itself become a security risk, giving attackers additional time to disrupt systems and services.

Several important challenges still remain in current IDS research. First, many deep learning-based IDS models operate as black-box systems that provide only prediction labels and confidence scores without explaining how the decision was made \cite{neupane2022explainable}. This makes it difficult for security analysts to fully trust or interpret the model outputs. Second, Explainable Artificial Intelligence (XAI) \cite{gunning2017explainable} techniques such as SHAP \cite{lundberg2017unified} and LIME \cite{ribeiro2016lime} can identify which features influenced a prediction, but they still do not provide practical mitigation guidance for responding to the attack. Third, valuable cybersecurity knowledge sources such as NIST \cite{moller2023nist} guidelines, MITRE ATT\&CK \cite{MITRE2025Enterprise}, and ENISA \cite{ipekinstitutional} recommendations contain detailed mitigation strategies, but these documents are extensive and difficult to consult during real-time incident response.

To address these limitations, this paper proposes a modular end-to-end framework that combines deep learning-based intrusion detection, and RAG for both attack detection and mitigation generation. The framework not only identifies cyber-attacks with high accuracy, but also generates structured and evidence-based mitigation reports using authoritative cybersecurity knowledge sources. The main contributions of this work are summarized as follows:

\begin{enumerate}
    \item We design a structured prompting strategy that combines the predicted attack class, important network features, flow metadata, and security interpretations to provide context-aware inputs for the language model. This helps the model generate mitigation guidance based on the actual attack evidence detected by the IDS.
    
    \item  We develop a hybrid retrieval pipeline that combines BM25 \cite{robertson2009probabilistic} keyword search, FAISS-based \cite{Johnsonetal2017} dense vector retrieval, and cross-encoder reranking to retrieve relevant information from a cybersecurity knowledge base built from authoritative sources including NIST \cite{moller2023nist} and MITRE \cite{MITRE2025Enterprise} documents.
    
    \item  Using the retrieved cybersecurity knowledge, the framework generates structured mitigation reports with supporting citations from trusted sources. The generated reports provide actionable recommendations for security analysts during incident response.
    
    \item  We evaluate the quality of the generated mitigation reports against expert-written ground truth explanations using BERTScore, ROUGE, and BLEU metrics. Experimental results show that the RAG-enhanced framework produces more accurate and reliable outputs compared to standalone LLM-based generation.
\end{enumerate}

The remainder of this paper is organized as follows. Section~\ref{sec:related} surveys related work and identifies the specific gaps this paper addresses. Section~\ref{sec:method} presents the full methodology, including datasets, DNN architectures, prompt construction, knowledge base design, retrieval architecture, and LLM configuration. Section~\ref{sec:results} reports and discusses experimental results. Section~\ref{sec:conclusion} concludes and outlines future work.

\section{Related Work}
\label{sec:related}
Recent advancements in cybersecurity have significantly improved the performance of IDS through the use of machine learning and deep learning models. Traditional IDS approaches mostly depended on signature-based or rule-based techniques, which often struggle to detect previously unseen attacks and complex traffic patterns \cite{lippmann2000evaluating}. To overcome these limitations, researchers introduced deep learning architectures capable of learning hidden representations directly from network traffic data.

\begin{table*}[t]
\centering
\caption{Comparison of Existing IDS, XAI, and RAG-Based Cybersecurity Systems}
\label{tab:related_work_comparison}

\footnotesize
\setlength{\tabcolsep}{4pt}
\renewcommand{\arraystretch}{1.15}

\resizebox{\textwidth}{!}{

\begin{tabular}{|p{3.5cm}|c|c|c|c|c|c|c|}
\hline

\textbf{Work} &
\textbf{DL-Based IDS} &
\textbf{XAI} &
\textbf{LLM/RAG} &
\textbf{Trusted KB} &
\textbf{Mitigation} &
\textbf{Hybrid Retrieval} &
\textbf{Detection-to-Mitigation} \\

\hline

Lippmann et al. \cite{lippmann2000evaluating}
& \xmark & \xmark & \xmark & \xmark & \xmark & \xmark & \xmark \\

\hline

Kim et al. \cite{kim2016lstm}
& \cmark & \xmark & \xmark & \xmark & \xmark & \xmark & \xmark \\

\hline

Yin et al. \cite{yin2017deep}
& \cmark & \xmark & \xmark & \xmark & \xmark & \xmark & \xmark \\

\hline

Shone et al. \cite{shone2018deep}
& \cmark & \xmark & \xmark & \xmark & \xmark & \xmark & \xmark \\

\hline

Vinayakumar et al. \cite{vinayakumar2019deep}
& \cmark & \xmark & \xmark & \xmark & \xmark & \xmark & \xmark \\

\hline

Lansky et al. \cite{lansky2021deep}
& \cmark & \xmark & \xmark & \xmark & \xmark & \xmark & \xmark \\

\hline

Lundberg and Lee \cite{lundberg2017unified}
& \xmark & \cmark & \xmark & \xmark & \xmark & \xmark & \xmark \\

\hline

Ribeiro et al. \cite{ribeiro2016lime}
& \xmark & \cmark & \xmark & \xmark & \xmark & \xmark & \xmark \\

\hline

Ferrag et al. \cite{ferrag2024revolutionizing}
& \cmark & Partial & \cmark & \xmark & \xmark & \xmark & \xmark \\

\hline

SecBERT \cite{huang2024secbert}   
& Partial & \xmark & \cmark & \xmark & \xmark & \xmark & \xmark \\

\hline

CyBERT \cite{ranade2021cybert}
& Partial & \xmark & \cmark & \xmark & \xmark & \xmark & \xmark \\

\hline

MoRSE \cite{simoni2025morse}
& \xmark & \xmark & \cmark & \cmark & Advisory & Partial & \xmark \\

\hline

CyberRAG \cite{blefari2025cyberrag}
& Partial & \cmark & \cmark & Partial & \cmark & Partial & Partial \\

\hline

Setiawan \& Soewito \cite{setiawan2025instruction}
& \xmark & \xmark & \cmark & \cmark & Partial & \xmark & \xmark \\

\hline

Pinto et al. \cite{pinto2023review}
& Partial & \xmark & \xmark & \xmark & Limited & \xmark & \xmark \\

\hline

Loi et al. \cite{loi2025shap}
& \cmark & \cmark & \xmark & \xmark & \xmark & \xmark & \xmark \\

\hline

Tjhai et al. \cite{tjhai2022xai}
& \cmark & \cmark & \xmark & \xmark & \xmark & \xmark & \xmark \\

\hline

\textbf{Proposed Framework}
& \cmark & \cmark & \cmark & \cmark & \cmark & \cmark & \cmark \\

\hline

\end{tabular}
}

\end{table*}

\subsection{Deep Learning-Based Intrusion Detection Systems}
Deep learning models such as Convolutional Neural Networks (CNNs) \cite{lecun1998convolutional}, Recurrent Neural Networks (RNNs) \cite{rumelhart1986learning}, Long Short-Term Memory (LSTM) networks \cite{hochreiter1997long}, and Autoencoders \cite{rumelhart1986learning} have shown strong performance in intrusion detection tasks \cite{lansky2021deep}. These models can automatically learn complex spatial and temporal relationships from high-dimensional network traffic data, making them more effective than traditional machine learning approaches.

Kim et al. \cite{kim2016lstm} introduced an LSTM-based IDS framework capable of learning sequential traffic behavior for anomaly detection. Their work demonstrated that recurrent architectures can effectively capture dependencies in network flows. Similarly, Yin et al. \cite{yin2017deep} evaluated RNN-based IDS models and showed significant improvements over traditional classifiers such as Support Vector Machines (SVMs) \cite{cortes1995support} and Random Forests \cite{breiman2001random}. Their study highlighted the importance of feature learning in network intrusion detection.

Shone et al. \cite{shone2018deep} proposed a non-symmetric deep autoencoder architecture combined with Random Forest \cite{breiman2001random} classification for feature extraction and intrusion detection. Their work demonstrated that unsupervised feature learning can improve detection performance while reducing feature engineering complexity. Vinayakumar et al. \cite{vinayakumar2019deep} later presented a comprehensive deep learning framework for cyber threat detection using datasets such as UNSW-NB15 \cite{MoustafaSlay2015} and CICIDS2017 \cite{Sharafaldinetal2018}. Their results showed that deep neural networks can achieve high classification accuracy under complex attack scenarios.

Although these approaches achieved impressive detection performance, most of them focused mainly on attack classification accuracy. They generally lacked explainability, contextual reasoning, and actionable mitigation support for security analysts. 

\subsection{Explainable Artificial Intelligence in IDS}
As deep learning models became more complex, researchers started investigating XAI techniques to improve trust and interpretability in IDS. Security analysts often require explanations regarding why a network flow was classified as malicious before taking operational decisions.

Lundberg and Lee \cite{lundberg2017unified} introduced SHAP (SHapley Additive exPlanations), which provides feature-level importance scores for machine learning predictions. Ribeiro et al. \cite{ribeiro2016lime} proposed LIME (Local Interpretable Model-Agnostic Explanations), a model-agnostic explanation framework capable of interpreting local predictions. Both SHAP and LIME have been widely applied in cybersecurity research to improve IDS interpretability.

Several recent IDS studies integrated explainability mechanisms into deep learning pipelines. For example, Tjhai et al. \cite{tjhai2022xai} used SHAP explanations to identify influential network traffic features contributing to attack detection decisions. Similarly, Loi et al. \cite{loi2025shap} demonstrated that explainable IDS frameworks can significantly improve analyst understanding and trust during threat investigation.

Despite these improvements, most explainable IDS systems still focus only on feature attribution and attack interpretation. They generally do not provide contextual mitigation recommendations or operational guidance after detection.

\subsection{Transformer and LLM-Based Cybersecurity Systems}
The success of transformer architectures and LLMs has influenced cybersecurity research. Transformer-based models can capture contextual relationships better in sequential data compared to traditional CNN and RNN architectures \cite{vinayakumar2019deep}.

Ferrag et al. \cite{ferrag2024revolutionizing} introduced SecurityBERT, a lightweight transformer-based intrusion detection model designed for IoT and IIoT environments. Their framework used Privacy-Preserving Fixed Length Encoding (PPFLE) to convert network traffic into a textual representation suitable for transformer processing while preserving sensitive information. 

Other cybersecurity-focused language models such as SecBERT \cite{huang2024secbert} and CyBERT \cite{ranade2021cybert} further demonstrated the effectiveness of transformer architectures for threat analysis, malware detection, and vulnerability understanding. However, many transformer-based IDS systems still operate mainly as classification frameworks. While they improve contextual understanding, they often lack reliable retrieval mechanisms and grounded mitigation generation pipelines \cite{sandaruwan2024leveraging}.

\subsection{RAG in Cybersecurity}
RAG has emerged as a promising approach for improving the reliability and factual grounding of LLMs \cite{lewis2020retrieval}. Instead of relying only on the model’s pre-trained knowledge, RAG systems retrieve relevant external documents during response generation.

In cybersecurity, RAG has been utilized to reduce hallucinations and improve the accuracy of generated security recommendations. Simoni et al. \cite{simoni2025morse} introduced MoRSE, a chatbot that combines structured and unstructured retrieval pipelines to answer cybersecurity-related questions. Their results showed that retrieval-enhanced responses were significantly more accurate and context-aware than standard LLM outputs.

Blefari et al. \cite{blefari2025cyberrag} proposed CyberRAG, an agent-based RAG framework that combines multiple attack-specific classifiers with retrieval-enhanced explanation generation. Their system achieved strong attack reporting performance while improving interpretability and operational usability.

Setiawan and Soewito \cite{setiawan2025instruction} explored a multi-agent RAG framework for automatic Snort rule generation using cybersecurity knowledge sources such as Common Weakness Enumeration (CWE). Their system demonstrated that instruction-based prompting combined with retrieval mechanisms can improve threat response automation.

\subsection{Automated Threat Mitigation and Incident Response}
Automated mitigation and incident response systems have also got the attention in recent years due to the increasing volume of security alerts generated by modern networks. Traditional Security Orchestration, Automation, and Response (SOAR) platforms help automate incident handling workflows but often depend heavily on manually crafted rules and predefined rule-books.

Pinto et al. \cite{pinto2023review} discussed the growing cybersecurity risks in smart grid infrastructures and highlighted the limitations of current detection and mitigation frameworks under real-world attack scenarios. Their work emphasized the importance of integrating intelligent detection systems with adaptive response mechanisms.

Recent studies also explored the integration of cybersecurity knowledge bases such as NIST \cite{moller2023nist}, MITRE ATT\&CK \cite{MITRE2025Enterprise}, ENISA \cite{ENISA2022SOC} into automated incident response systems. However, most existing automated mitigation frameworks either operate independently from IDS systems or rely heavily on generic responses without incorporating contextual attack evidence derived from network traffic analysis.

\subsection{Research Gap and Motivation}

When analyzing existing literature, several important limitations become clear. Deep learning-based IDS frameworks achieve strong attack detection performance but usually lack explainability and mitigation support. Explainable IDS systems improve interpretability but generally stop at feature attribution without generating actionable responses. Transformer and LLM-based cybersecurity systems improve contextual reasoning but often lack authoritative grounding and retrieval validation. Similarly, many RAG-based cybersecurity frameworks focus on threat  retrieval rather than tightly integrating intrusion detection with mitigation generation.

Another important limitation is that most existing works do not address real-world operational challenges such as confidence-aware analysis, hybrid retrieval architectures, reranking mechanisms, and grounded mitigation reporting. Very few studies provide an end-to-end framework that combines intrusion detection, explainability, authoritative retrieval, and context-aware mitigation generation in a unified pipeline.

Motivated by these limitations, this paper proposes a complete IDS framework that combines deep learning-based intrusion detection with a RAG pipeline for mitigation report generation. Unlike existing approaches, the proposed framework integrates explanation-aware prompt construction, hybrid lexical-semantic retrieval, cross-encoder reranking, confidence-aware classification, and grounded mitigation generation using trusted cybersecurity sources. The system not only detects attacks but also generates structured, actionable, and citation-grounded mitigation reports suitable for real-world cybersecurity operations.

\begin{figure*}
    \centering
    \includegraphics[width=\textwidth]{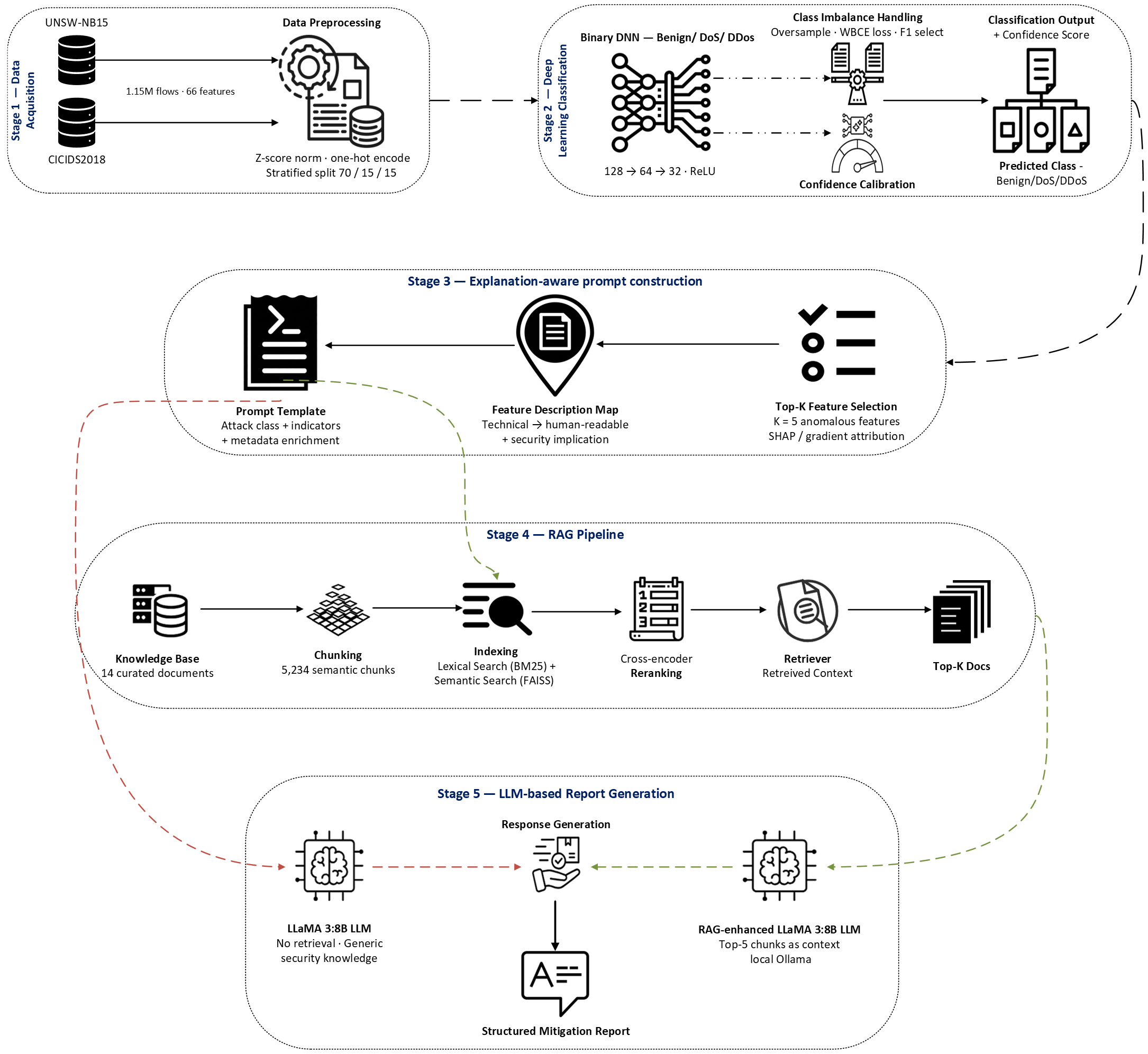}
    \caption{Proposed end-to-end framework for intrusion detection and RAG-enhanced mitigation generation. The architecture integrates DNN-based attack classification, explanation-aware prompt construction, hybrid retrieval using BM25 and FAISS with cross-encoder reranking, and LLM-based mitigation report generation using authoritative cybersecurity knowledge sources.}
    \label{fig:proposed_framework}
\end{figure*}

\section{Proposed Methodology}
\label{sec:method}

This section presents the complete technical design of the proposed IDS--RAG framework. The system follows a five-stage modular pipeline: (1) data acquisition and preprocessing, (2) confidence-calibrated DNN classification, (3) explanation-aware prompt construction, (4) hybrid RAG retrieval and reranking, and (5) structured LLM report generation. The overall architecture is illustrated in Fig.~\ref{fig:proposed_framework}.

\subsection{System Architecture Overview}

The proposed pipeline ingests raw network flow records, preprocesses them into a unified 66-dimensional feature space, classifies each flow using an ensemble of three binary DNNs, extracts the most anomalous features for prompt construction, retrieves relevant cybersecurity guidance through a hybrid retrieval system, and generates a structured mitigation report using a locally deployed LLM. Each stage is independently modular: the classifier can be retrained on new traffic data without modifying the RAG system, and the knowledge base can be extended with new authorative sources without retraining the classifier.

\subsection{Stage 1: Data Acquisition and Preprocessing}

We have used two publicly available benchmark datasets. The CICIDS2018 dataset \cite{Sharafaldinetal2018} provides 1.15 million flow records spanning three classes (Benign, DoS, DDoS) across 66 flow-level features. The UNSW-NB15 dataset~\cite{moustafa2015unsw} provides a complementary evaluation environment with 49 original features across ten classes. To enable cross-dataset evaluation, all nine attack classes in UNSW-NB15 are remapped to a three-class scheme: complex multi-stage attacks (Fuzzers, Exploits, Generic, Reconnaissance) are combined under DDoS due to their distributed, high-impact characteristics, while direct volumetric attacks are assigned to DoS.

Numeric features such as packet counts, byte counts, TTL values etc. are retained as-is. Categorical features like protocol, service, connection state etc. are one-hot encoded. Zero-padding features are constructed to bridge the 49-to-66 dimensional gap, ensuring that both datasets share an identical 66-dimensional feature space. A feature interpretation dictionary is maintained that maps each technical feature name to a human-readable description and a security implication string. Table~\ref{tab:feature_map} shows ten representative entries from this dictionary.

\begin{table}
\centering
\caption{Representative Feature Interpretation Dictionary Entries}
\label{tab:feature_map}
\small
\renewcommand{\arraystretch}{1.2}
\begin{tabular}{p{1.7cm}p{2.6cm}p{2.8cm}}
\toprule
\textbf{Feature} & \textbf{Description} & \textbf{Security Implication} \\
\midrule
sttl & Source Time-to-Live & Irregular values may indicate spoofed packets \\
ct\_stat\_ttl & Conn. state transitions & Abnormal state cycles signal protocol abuse \\
ct\_srv\_dst & Conn. to same service & Repeated targeting of one endpoint \\
sload & Source byte throughput & High values consistent with flooding \\
dload & Destination byte load & Asymmetric loads suggest amplification \\
Flow Bytes/s & Bytes per second & High rate indicates volumetric activity \\
Flow Packets/s & Packets per second & Scan DoS behaviour at high values \\
Init Win Bytes & Initial TCP window & Anomalous values flag scan tools \\
Avg Packet Size & Mean payload bytes & Small uniform sizes indicate flood \\
Protocol & L4 protocol number & UDP/ICMP often used in amplification \\
\bottomrule
\end{tabular}
\end{table}

A scikit-learn~\cite{garreta2017scikit} pipeline applies $z$-score normalization to all numeric features:
\begin{equation}
\hat{x}_j = \frac{x_j - \mu_j}{\sigma_j}
\label{eq:zscore}
\end{equation}
where $\mu_j$ and $\sigma_j$ are the mean and standard deviation of feature $j$ computed over the training split. Stratified 70/15/15 train/validation/test splits are used across both datasets to preserve class proportions and ensure unbiased evaluation under severe class imbalance.

\subsection{Stage 2: DNN Classification}

Rather than training a single three-way classifier, we deploy an ensemble of three independently trained binary DNNs \cite{yuan2023comprehensive}, one per class Benign-vs-Rest, DoS-vs-Rest, DDoS-vs-Rest. This one-vs-rest strategy facilitates the learning objective of each model by allowing it to focus on a single decision boundary rather than simultaneously separating multiple attack categories.

\subsubsection{Network Architectures}
Two different DNN architectures are employed depending on the distribution of the dataset.

For the CICIDS2018 \cite{Sharafaldinetal2018} dataset, the model consists of three fully connected hidden layers with dimensions $[128, 64, 32]$. Each hidden layer uses the ReLU activation function followed by dropout regularization \cite{zhang2018multiple} with a dropout probability of $p = 0.3$ to reduce overfitting and improve generalization.

For the UNSW-NB15 \cite{moustafa2015unsw} dataset, the model contains four fully connected hidden layers with dimensions $[256, 128, 64, 32]$. ReLU activation is applied after each layer, while batch normalization \cite{jung2019restructuring} is incorporated to stabilize training and reduce internal covariate shift. In addition, graduated dropout regularization is used with probabilities decreasing from $0.4$ to $0.2$ across the deeper layers.

Both architectures terminate with a softmax output layer~\cite{liang2017soft}. The full forward pass for an input flow $\mathbf{x} \in \mathbb{R}^{66}$ through the $k$-layer network is:

\begin{equation}
    \mathbf{h}^{(0)} = \mathbf{x} 
\end{equation}
\begin{equation}
\mathbf{h}^{(l)} = \mathrm{ReLU}\!\left(\mathrm{BN}\!\left(\mathbf{W}^{(l)} \mathbf{h}^{(l-1)} + \mathbf{b}^{(l)}\right)\right), \; l = 1, \ldots, k-1 
\end{equation}
\begin{equation}
\hat{y} = \mathrm{Softmax}\!\left(\mathbf{W}^{(k)} \mathbf{h}^{(k-1)} + \mathbf{b}^{(k)}\right) 
\end{equation}

where $\mathbf{W}^{(l)}$ and $\mathbf{b}^{(l)}$ represent the learnable weights and bias parameters of layer $l$, and BN denotes the batch normalization operation, which is applied only in the UNSW-NB15 architecture.

\subsubsection{Class Imbalance Problem}
In the UNSW-NB15\cite{moustafa2015unsw} dataset, DoS traffic constitutes only 0.64\% of total samples. Under standard cross-entropy training, this causes the optimizer to achieve a slightly low loss by predicting the majority class for nearly all inputs, yielding DoS recall close to 0.

\subsubsection{Class-Weighted Binary Cross-Entropy}
To address this, the binary cross-entropy loss is modified with inverse-frequency class weights. For a single sample with true label $y \in \{0,1\}$ and predicted probability $\hat{y}$, the weighted loss is:
\begin{equation}
\mathcal{L}_{\text{WBCE}}(y, \hat{y}) = -\bigl[w_1 \, y \log(\hat{y}) + w_0 \, (1-y) \log(1-\hat{y})\bigr]
\label{eq:wbce}
\end{equation}
where the class weights $w_c$ are computed as:
\begin{equation}
w_c = \frac{N}{2 \, N_c}
\label{eq:wc}
\end{equation}
$N$ is the total number of training samples and $N_c$ is the number of samples in class $c$. The total training objective over a minibatch of $N$ samples is:
\begin{equation}
\mathcal{L}_{\text{total}} = \frac{1}{N} \sum_{i=1}^{N} \mathcal{L}_{\text{WBCE}}(y_i, \hat{y}_i)
\label{eq:total_loss}
\end{equation}
This formulation implements empirical risk minimization under class-balanced constraints, ensuring that minority-class errors receive proportionally larger gradient contributions~\cite{xu2020class}.

\subsubsection{Additional Balancing Interventions}
Beyond loss weighting, three further strategies are applied specifically for the UNSW-NB15 DoS classifier: (i) random oversampling of the minority DoS class to a 1:5 ratio relative to the Benign class~\cite{mohammed2020machine}, (ii) controlled undersampling of the Benign majority, and (iii) model selection based on macro-averaged F1-score rather than accuracy. The combined interventions raise DoS recall from 0.30\% to 97.6\%---a 325-fold improvement (Table~\ref{tab:balancing_ablation}).

\begin{table}
\centering
\caption{Ablation Study: Effect of Each Balancing Intervention on UNSW-NB15 DoS Recall}
\label{tab:balancing_ablation}
\small
\renewcommand{\arraystretch}{1.2}
\begin{tabular}{lc}
\toprule
\textbf{Configuration} & \textbf{DoS Recall (\%)} \\
\midrule
Baseline (no balancing)                          & 0.30 \\
+ Class-weighted BCE                              & 28.4 \\
+ Random oversampling (1:5)                      & 71.2 \\
+ Controlled undersampling                       & 84.9 \\
+ F1-based model selection (\textbf{Full System}) & \textbf{97.6} \\
\bottomrule
\end{tabular}
\end{table}

\subsubsection{Ensemble Prediction and Confidence Calibration}
All three classifiers receive the same input flow simultaneously. Let $p_c$ denote the softmax probability assigned by classifier $c \in \{\mathrm{Benign}, \mathrm{DoS}, \mathrm{DDoS}\}$. The final prediction and confidence are given by:
\begin{equation}
\hat{c} = \arg\max_{c} \; p_c, \quad \mathrm{Confidence} = \max_{c} \; p_c
\label{eq:ensemble}
\end{equation}

The scalar confidence score is mapped to a discrete tier for SOC prioritization, as shown in Table~\ref{tab:confidence_tiers}. The full procedure is presented as Algorithm~\ref{alg:ensemble}.

\begin{table}[!t]
\centering
\caption{Confidence Tier Mapping for SOC Triage}
\label{tab:confidence_tiers}
\small
\renewcommand{\arraystretch}{1.2}
\begin{tabular}{lll}
\toprule
\textbf{Confidence} & \textbf{Tier} & \textbf{SOC Action} \\
\midrule
$\geq 0.95$           & Very High & Fully automated response \\
$[0.70, 0.95)$        & High      & Analyst review within 15 min \\
$[0.50, 0.70)$        & Medium    & Analyst review within 1 hr \\
$< 0.50$              & Low       & Manual deep inspection \\
\bottomrule
\end{tabular}
\end{table}

\begin{algorithm}
\caption{Ensemble Confidence-Calibrated Classification}
\label{alg:ensemble}
\begin{algorithmic}[1]
\REQUIRE Feature vector $\mathbf{x} \in \mathbb{R}^{66}$; trained classifiers $f_B, f_D, f_{DD}$
\ENSURE Predicted class $\hat{c}$, confidence $\mathrm{conf}$, tier label $t$
\STATE $p_B  \gets f_B(\mathbf{x})[1]$ \COMMENT{Benign probability}
\STATE $p_D  \gets f_D(\mathbf{x})[1]$ \COMMENT{DoS probability}
\STATE $p_{DD} \gets f_{DD}(\mathbf{x})[1]$ \COMMENT{DDoS probability}
\STATE $\hat{c}  \gets \arg\max\{p_B, p_D, p_{DD}\}$
\STATE $\mathrm{conf} \gets \max\{p_B, p_D, p_{DD}\}$
\IF{$\mathrm{conf} \geq 0.95$}
    \STATE $t \gets \textsc{Very High}$
\ELSIF{$\mathrm{conf} \geq 0.70$}
    \STATE $t \gets \textsc{High}$
\ELSIF{$\mathrm{conf} \geq 0.50$}
    \STATE $t \gets \textsc{Medium}$
\ELSE
    \STATE $t \gets \textsc{Low}$
\ENDIF
\RETURN $\hat{c}, \mathrm{conf}, t$
\end{algorithmic}
\end{algorithm}

All models are trained with the Adam \cite{kingma2014adam} optimizer, an initial learning rate of $1 \times 10^{-3}$, batch size 512, and early stopping with patience of 5 epochs monitored on validation F1-score. Training converges within 14--19 epochs across all classifiers on both datasets.

\subsection{Stage 3: Explanation-Aware Prompt Construction}

Conventional approaches query an LLM with a generic question such as ``Describe mitigation strategies for a DDoS attack'' which produces textbook-level advice contextually detached from the specific event. Our approach constructs a structured, evidence-rich prompt that binds the LLM's reasoning to the concrete evidence produced by the classifier.

\subsubsection{Top-$K$ Feature Selection}
Following classification, the $K = 5$ features most influential to the model's prediction are extracted. Feature importance scores are computed using gradient-based attribution~\cite{ancona2019gradient}. The gradient of the predicted class score with respect to each input dimension is computed and its absolute value taken as a proxy for feature influence:

\begin{equation}
\mathrm{Importance}(j) = \left| \frac{\partial \hat{y}_{\hat{c}}}{\partial x_j} \right|
\label{eq:importance}
\end{equation}

Features are sorted in descending order of importance and the top-5 selected. For the experiments in this paper, five security-relevant features per dataset were additionally identified through domain knowledge:

\begin{itemize}
    \item \textbf{CICIDS2018:} Flow Bytes/s, Flow Packets/s, Init Win Bytes Forward, Protocol, Average Packet Size.
    \item \textbf{UNSW-NB15:} \texttt{sttl} (source TTL), \texttt{ct\_state\_ttl} (connection state transitions), \texttt{dload} (destination byte load), \texttt{sload} (source throughput), \texttt{ct\_srv\_dst} (repeated service connections).
\end{itemize}

\subsubsection{Feature Description Mapping}
Each selected feature is mapped through the interpretation dictionary (Table~\ref{tab:feature_map}) to produce a natural-language evidence string. For example, a flow with $\texttt{sttl} = 245$ becomes \emph{``Source TTL of 245 is elevated and irregular; may indicate packet spoofing or TTL manipulation consistent with DoS traffic engineering,''} and $\text{Flow Bytes/s} = 4.7 \times 10^6$ becomes \emph{``Abnormally high byte rate suggests sustained bulk data flooding, consistent with volumetric denial-of-service activity.''}

\subsubsection{Prompt Template Construction}
The structured prompt is assembled using LangChain's \texttt{PromptTemplate}
framework~\cite{liu2023prompting} and comprises four semantically distinct
blocks, each contributing different contextual information to guide report
generation. Table~\ref{tab:prompt_blocks} summarizes the four blocks, their
constituent variables, and the pipeline stage responsible for populating each.
The complete template is shown in Box~\ref{box:prompt}.

\begin{table}
\centering
\caption{Prompt Block Structure and Variable Sources}
\label{tab:prompt_blocks}
\renewcommand{\arraystretch}{1.3}
\begin{tabular}{p{2.0cm} p{3.2cm} p{2.2cm}}
\hline
\textbf{Block} & \textbf{Variables} & \textbf{Source Stage} \\
\hline
Detection Context
  & \texttt{attack\_class},
    \texttt{confidence\_score},
    \texttt{confidence\_tier},
    \texttt{dataset}
  & Stage~2 (DNN) \\
\hline
Flow Metadata
  & \texttt{flow\_id},
    \texttt{timestamp},
    \texttt{src\_ip},
    \texttt{src\_port},
    \texttt{dst\_ip},
    \texttt{dst\_port},
    \texttt{protocol}
  & Stage~1 (preprocessing) \\
\hline
Anomalous Indicators
  & \texttt{feature\_\{1..K\}\_name},
    \texttt{feature\_\{1..K\}\_value},
    \texttt{feature\_\{1..K\}\_interpretation}
  & Stage~3 (XAI) \\
\hline
Retrieved Knowledge
  & \texttt{retrieved\_chunks}
  & Stage~4 (RAG) \\
\hline
\end{tabular}
\end{table}

The system role instruction fixes the LLM persona as a senior cybersecurity
analyst and imposes an explicit citation requirement for NIST Special
Publications and MITRE ATT\&CK techniques, ensuring that generated reports
remain grounded in authoritative sources rather than relying on the model's
parametric knowledge alone. The detection context block provides the predicted
attack class, scalar confidence score, and SOC triage tier from
Algorithm~\ref{alg:ensemble}, binding the model's reasoning to the specific
classification outcome. The flow metadata block supplies network-level
identifiers that enable traceability in automated reporting pipelines and
allow the LLM to tailor its language to the operational context of the event.
The anomalous indicators block injects the top-$K$ feature--interpretation
pairs produced in Stage~3, providing the concrete evidence from which the
model constructs its rationale. The retrieved knowledge block is populated
by the RAG pipeline with the top-$k = 5$ chunks selected by the cross-encoder;
in the vanilla LLM baseline condition this field is left empty, isolating
the contribution of retrieval augmentation. Finally, the report request
block specifies the five-section output structure required of the model.


\begin{figure}[!t]
\refstepcounter{boxcounter}
\label{box:prompt}
\begin{tcolorbox}[
    enhanced,
    breakable,
    colback       = gray!4,
    colframe      = black!75,
    boxrule       = 0.6pt,
    arc           = 2pt,
    left          = 6pt,
    right         = 6pt,
    top           = 4pt,
    bottom        = 4pt,
    title         = {Explanation-Aware Prompt Template},
    fonttitle     = \bfseries\small,
    coltitle      = white,
    colbacktitle  = black!80,
    attach boxed title to top left = {yshift=-2pt, xshift=4pt},
    boxed title style = {arc=1.5pt, boxrule=0pt}
]
\ttfamily\footnotesize
\textbf{SYSTEM:}\\
\hspace*{1em} You are a senior cybersecurity analyst.\\
\hspace*{1em} Generate a structured incident response report\\
\hspace*{1em} based on the evidence provided. Cite NIST Special\\
\hspace*{1em} Publications and MITRE ATT\&CK techniques explicitly\\
\hspace*{1em} where applicable.\\[4pt]
\textbf{USER:}\\[2pt]
\textcolor{black!55}{\textit{[DETECTION CONTEXT]}}\\
\hspace*{1em} Attack class \hfill \texttt{\{attack\_class\}}\\
\hspace*{1em} Confidence   \hfill \texttt{\{confidence\_score\} (\{confidence\_tier\})}\\
\hspace*{1em} Dataset      \hfill \texttt{\{dataset\}}\\[4pt]
\textcolor{black!55}{\textit{[NETWORK FLOW METADATA]}}\\
\hspace*{1em} Flow ID      \hfill \texttt{\{flow\_id\}}\\
\hspace*{1em} Timestamp    \hfill \texttt{\{timestamp\}}\\
\hspace*{1em} Source       \hfill \texttt{\{src\_ip\}:\{src\_port\}}\\
\hspace*{1em} Destination  \hfill \texttt{\{dst\_ip\}:\{dst\_port\}}\\
\hspace*{1em} Protocol     \hfill \texttt{\{protocol\}}\\[4pt]
\textcolor{black!55}{\textit{[KEY ANOMALOUS INDICATORS]}}\\
\hspace*{1em} Top-\texttt{\{K\}} features most influential in this decision:\\
\hspace*{2em} Feature 1 :\ \texttt{\{feature\_1\_name\}} $=$ \texttt{\{feature\_1\_value\}}\\
\hspace*{3em} $\hookrightarrow$\ \texttt{\{feature\_1\_interpretation\}}\\
\hspace*{2em} $\vdots$\\
\hspace*{2em} Feature K :\ \texttt{\{feature\_K\_name\}} $=$ \texttt{\{feature\_K\_value\}}\\
\hspace*{3em} $\hookrightarrow$\ \texttt{\{feature\_K\_interpretation\}}\\[4pt]
\textcolor{black!55}{\textit{[RETRIEVED KNOWLEDGE CONTEXT]}}\\
\hspace*{1em} \texttt{\{retrieved\_chunks\}} \hfill
              \textcolor{black!50}{$\triangleright$ injected by RAG pipeline}\\[4pt]
\textcolor{black!55}{\textit{[REPORT REQUEST]}}\\
\hspace*{1em} Generate a structured report with sections:\\[2pt]
\hspace*{2em}
\begin{tabular}{@{}ll@{}}
  1.\ Rationale            & 4.\ Threat Assessment \\
  2.\ Key Indicators       & 5.\ Recommendations   \\
  3.\ Confidence Assessment &                      \\
\end{tabular}
\end{tcolorbox}
\end{figure}
\subsubsection{Metadata Enrichment}
The prompt incorporates flow metadata~\cite{peel2017ground} (IP addresses, ports, protocol, timestamp, dataset provenance, confidence tier) to enable traceability in automated reporting systems and to allow the LLM to tailor its language to the specific operational context.

\subsection{Stage 4: RAG Pipeline}

\subsubsection{Comprehensive Knowledge Base Construction}
The RAG system's retrieval quality depends on the quality of the knowledge base it searches. We construct a domain-specific, authoritative knowledge base from fourteen carefully selected cybersecurity documents. Nine NIST Special Publications~\cite{moller2023nist} were included: SP~800-61 Rev.~2 (incident handling), SP~800-94 (IDS guidelines), SP~800-83 Rev.~1 (malware incident response), SP~800-115 (security testing), SP~800-137 (continuous monitoring), SP~800-92 (log management), SP~800-123, SP~800-125, and SP~800-144. The MITRE ATT\&CK Enterprise Framework~\cite{MITRE2025Enterprise} was included with particular emphasis on T1498 (Network Denial of Service) and T1498.001 (Direct Network Flood). Additional sources include the CIS Critical Security Controls, SOC incident response playbooks, and curated attack signature databases~\cite{alyousef2019dynamically}.

\subsubsection{Semantic Chunking}
Documents are segmented using a semantic chunking strategy~\cite{qu2025semantic} rather than naive fixed-size splitting. Chunk boundaries are placed at semantic units with a 200-token overlap between adjacent chunks, preserving conceptual continuity across boundaries. Each chunk is annotated with: (i) traffic relevance label (Benign / DoS / DDoS), (ii) source document identifier, (iii) NIST/MITRE reference, and (iv) section identifier. The final knowledge base contains 5{,}234 semantically coherent, metadata-tagged chunks.

\subsubsection{Hybrid Retrieval Architecture}
The retrieval system combines two complementary paradigms lexical and semantic search followed by a reranking stage.

\paragraph{Lexical Retrieval}
The lexical retrieval stage uses the BM25 ranking function \cite{robertson2009probabilistic} to measure the relevance between the user query $Q$ and each knowledge document $D$. The BM25 relevance score is computed as:

\begin{equation}
\small
\begin{aligned}
\mathrm{Score}_{\mathrm{BM25}}(D,Q)
&=
\sum_i \mathrm{IDF}(q_i)
\\
&\quad \times
\frac{
f(q_i,D)(k_1+1)
}{
f(q_i,D)
+
k_1\!\left(
1-b+b\frac{|D|}{\mathrm{avgdl}}
\right)
}
\end{aligned}
\label{eq:bm25}
\end{equation}

where $f(q_i,D)$ denotes the frequency of query term $q_i$ in document $D$, $|D|$ represents the document length, and $\mathrm{avgdl}$ is the average document length across the corpus. The parameter $k_1$ controls term frequency saturation, while $b$ determines the degree of document length normalization. In this paper, standard BM25 hyper-parameters $k_1 = 1.5$ and $b = 0.75$ are used.

\paragraph{Semantic Retrieval}
All knowledge chunks are encoded into 768-dimensional dense vectors using the \emph{all-mpnet-base-v2} sentence transformer~\cite{galli2024performance} and indexed in a FAISS~\cite{danopoulos2019approximate} approximate nearest-neighbor index. Cosine similarity~\cite{gunawan2018implementation} between query embedding $\mathbf{q}$ and chunk embedding $\mathbf{d}$ is:

\begin{equation}
\mathrm{Sim}(\mathbf{q}, \mathbf{d}) = \frac{\mathbf{q} \cdot \mathbf{d}}{\|\mathbf{q}\|_2 \, \|\mathbf{d}\|_2}
\label{eq:cosine}
\end{equation}

\paragraph{Query Expansion}
To improve retrieval coverage, the query is expanded using a cybersecurity-specific synonym dictionary that incorporates semantically related attack terminology. For example, the term ``DoS'' is expanded with related expressions such as ``denial of service,'' ``resource exhaustion,'' and ``flooding,'' while ``DDoS'' is associated with terms including ``distributed denial of service,'' ``botnet flood,'' and ``amplification attack.'' The expanded query representation is applied to both the lexical BM25 retrieval stage and the semantic embedding-based retrieval stage in order to improve recall and increase the possibility of retrieving contextually relevant cybersecurity knowledge.

\paragraph{Score Fusion}
Normalised BM25 and cosine similarity scores are combined as a weighted fusion:
\begin{equation}
\mathrm{Score}_{\text{fusion}}(D,Q) = 0.60 \cdot \mathrm{Sim}_{\text{sem}}(D,Q) + 0.40 \cdot \mathrm{Score}_{\text{BM25}}(D,Q)
\label{eq:fusion}
\end{equation}

The 60/40 weighting was determined empirically and favours semantic relevance while preserving sensitivity to exact technical terminology. The top 20--30 chunks are forwarded to reranking.

\paragraph{Cross-Encoder Reranking}
The \emph{ms-marco-MiniLM-L-6-v2} cross-encoder \cite{ravishankar2025novel} encodes each query--chunk pair, enabling fine-grained relevance interactions that bi-encoders cannot capture. Each of the 20--30 candidates receives a relevance score in $[0,1]$, and the top-$k = 5$ chunks are selected as final retrieval results. The complete pipeline is presented in Algorithm~\ref{alg:retrieval}.

\begin{algorithm}[!t]
\caption{Hybrid RAG Retrieval Pipeline}
\label{alg:retrieval}
\begin{algorithmic}[1]
\REQUIRE Query $q$ (attack class + features + metadata); knowledge base KB; $k = 5$
\ENSURE Top-$k$ ranked chunks $\mathcal{C}^{*}$
\STATE $q_{\text{exp}} \gets \mathrm{QueryExpand}(q, \mathrm{thesaurus})$
\STATE $\mathcal{C}_{\text{bm25}} \gets \mathrm{BM25}(q_{\text{exp}}, \mathrm{KB}, n{=}30)$
\STATE $\mathbf{q}_{\text{emb}}  \gets \mathrm{Encode}(q_{\text{exp}})$
\STATE $\mathcal{C}_{\text{sem}}  \gets \mathrm{FAISS}(\mathbf{q}_{\text{emb}}, \mathrm{KB}, n{=}30)$
\STATE $\mathcal{C}_{\text{cand}} \gets \mathrm{Dedup}(\mathcal{C}_{\text{bm25}} \cup \mathcal{C}_{\text{sem}})$
\FORALL{$d \in \mathcal{C}_{\text{cand}}$}
    \STATE $s_{\text{bm25}} \gets \mathrm{NormBM25}(d, q_{\text{exp}})$
    \STATE $s_{\text{sem}}  \gets \mathrm{Cos}(\mathbf{q}_{\text{emb}}, \mathrm{Encode}(d))$
    \STATE $s_{\text{fuse}} \gets 0.60 \cdot s_{\text{sem}} + 0.40 \cdot s_{\text{bm25}}$
\ENDFOR
\STATE $\mathcal{C}_{30} \gets \mathrm{Top\text{-}30}\{s_{\text{fuse}}\}$
\FORALL{$d \in \mathcal{C}_{30}$}
    \STATE $s_{\text{rr}} \gets \mathrm{CrossEnc}(q, d)$
\ENDFOR
\STATE $\mathcal{C}^{*} \gets \mathrm{Top\text{-}}k\{s_{\text{rr}}\}$
\RETURN $\mathcal{C}^{*}$
\end{algorithmic}
\end{algorithm}

\subsection{Stage 5: LLM-Based Report Generation}

LLaMA~3:8B~\cite{touvron2023llama}, accessed through the Ollama~\cite{marcondes2025using} framework for fully local inference, serves as the generation backbone. Network traffic data processed by an IDS is typically sensitive, and routing it to external API services creates both privacy and regulatory risks. 

\subsubsection{Structured Report Format}
The LLM is instructed via the prompt template (Listing~1) to produce a
five-section incident response report, each section serving a distinct
operational function for SOC analysts. The \textbf{Rationale} section links
the observed anomalous network features to the classified attack type, with
explicit citations to the applicable NIST Special Publication and MITRE
ATT\&CK technique. The \textbf{Key Indicators} section enumerates the
top-$K$ features selected in Stage~3 alongside their security
interpretations, providing the evidentiary basis for the classification.
The \textbf{Confidence Assessment} section evaluates detection reliability
by contextualizing the scalar confidence score within the SOC triage tier
produced by Algorithm~\ref{alg:ensemble}, allowing analysts to calibrate
their response urgency accordingly. The \textbf{Threat Assessment} section
characterizes the operational impact of the detected attack using the risk
assessment principles of NIST SP~800-30, including affected assets and
potential consequences. Finally, the \textbf{Recommendations} section
presents specific, prioritized mitigation actions drawn directly from the
top-$k$ retrieved knowledge chunks, each accompanied by its source
citation. The five-section structure is deliberately aligned with standard
SOC incident response workflows, ensuring that generated reports are
immediately actionable without requiring analyst reformatting or
supplementary research.

\subsubsection{Fallback Mechanism}
If retrieval returns fewer than three chunks with cross-encoder score above 0.5, a fallback knowledge context containing general NIST SP~800-61 incident response principles~\cite{bodeau2013cyber} is injected to guarantee consistent output quality.

\subsubsection{Vanilla LLM Baseline}
To isolate the contribution of RAG, a vanilla LLM baseline receives the identical explanation-aware prompt but with the \texttt{\{retrieved\_chunks\}} field replaced by an empty string. This controls for prompt design effects and isolates the value of retrieval augmentation.

\subsubsection{Expert Ground Truth Development}
To enable rigorous evaluation, expert-aligned ground truth explanations were constructed from primary sources rather than researcher-authored text. Ground truth texts were merged from NIST SP~800-61, NIST SP~800-94 \cite{NIST_SP800_53_R5_2020}, MITRE ATT\&CK T1498/T1498.001~\cite{MITRE2025Enterprise}, and RFC~4732~\cite{iab2006rfc}. Three ground truth documents were produced, one per class:
\begin{itemize}
    \item \textbf{Benign:} RFC-compliant protocol interactions, balanced bidirectional traffic, human-paced timing, correct TCP handshake sequences.
    \item \textbf{DoS:} single-source high-volume traffic, abnormal packet rates, resource exhaustion patterns, NIST SP~800-61 \cite{moller2023nist} incident handling procedures, MITRE T1498 \cite{MITRE2025Enterprise} indicators, rate limiting and SYN cookie recommendations.
    \item \textbf{DDoS:} distributed traffic sources, synchronized attack timing, botnet coordination, large-scale bandwidth consumption, DNS/NTP amplification, RFC~4732 \cite{iab2006rfc} mitigation guidance, MITRE T1498.001 \cite{MITRECWE2026} countermeasures.
\end{itemize}

Each ground truth document is 300--600 words, matching the target length of generated explanations to avoid artificial length-mismatch penalties.

\subsection{Implementation Details}
The model has been developed using Python 3.11 with PyTorch 2.1.0 for neural network training. Neural network training has been done on an amd Ryzen-5 3600X workstation with 32 GB RAM and NVIDIA RTX 2060 GPU running Windows 11. The RAG pipeline integrates FAISS \cite{danopoulos2019approximate} 1.7.4 for vector search, LangChain \cite{Chase_LangChain_2022} 0.1.0 for LLM management, and LLaMA 3:8B \cite{touvron2023llama} with 4-bit quantization using Ollama \cite{Ollama2024}. The model also uses OpenAI's text-embedding-ada-002\cite{openai_ada002} model for all embeddings.

\section{Result \& Discussion}
\label{sec:results}

This section evaluates the proposed IDS--RAG framework across both detection and mitigation tasks. The classification performance of the ensemble DNN is analyzed on the CICIDS2018 \cite{cicids2018} and UNSW-NB15 \cite{moustafa2015unsw} datasets, followed by an evaluation of the hybrid retrieval pipeline and the quality of the generated mitigation reports. Finally, qualitative analysis and operational feasibility are discussed to assess the framework’s practicality in real-world SOC environments.

\subsection{Classification Performance on CICIDS2018}
\begin{figure}[htbp]
    \centering
    \includegraphics[width=\columnwidth]{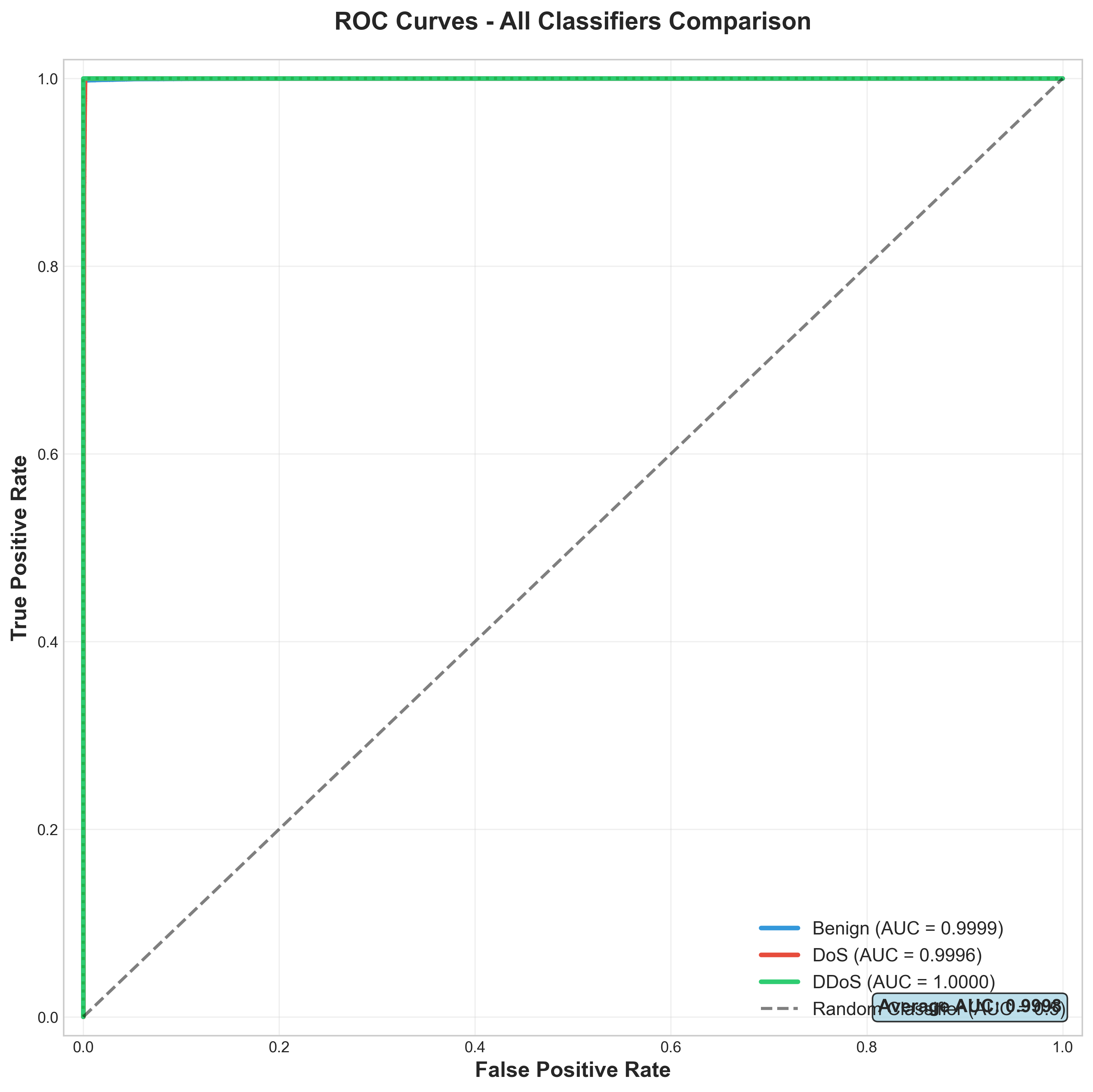}
    \caption{ROC Curves for CSECICIDS2018 Dataset}
    \label{fig:roc_cicids2018}
\end{figure}

Our proposed DNN architecture achieves an overall accuracy of \textbf{99.84\%} on the CICIDS2018 test set (250{,}050 flows). Table~\ref{tab:cic_perf} reports per-class precision, recall, and F1-score. DDoS detection is near-perfect ($F_1 = 0.9999$), with zero missed DDoS flows. DoS detection achieves $F_1 = 0.9935$. The only meaningful error source is 150 misclassified Benign flows out of 250{,}050 (0.06\% false-positive rate), well below thresholds that would cause alert fatigue in operational systems. ROC-AUC values are $\approx 1.0$ for all three classifiers (Fig.~\ref{fig:roc_cicids2018}), and confidence scores cluster tightly in $[0.98, 1.00]$ for correct predictions. 
\begin{table}
\centering
\caption{CICIDS2018 Classification Performance}
\label{tab:cic_perf}
\small
\renewcommand{\arraystretch}{1.2}
\begin{tabular}{lcccc}
\toprule
\textbf{Class} & \textbf{Precision} & \textbf{Recall} & \textbf{F1} & \textbf{Support} \\
\midrule
Benign        & 1.00 & 1.00 & 1.00 & 116{,}248 \\
DoS           & 0.99 & 1.00 & 0.99 &  58{,}342 \\
DDoS          & 1.00 & 1.00 & 1.00 &  75{,}460 \\
\midrule
Accuracy      & \multicolumn{4}{c}{1.00 \quad (250{,}050 samples)} \\
Macro Avg     & 1.00 & 1.00 & 1.00 & --- \\
Weighted Avg  & 1.00 & 1.00 & 1.00 & --- \\
\bottomrule
\end{tabular}
\end{table}

Training dynamics confirm robust generalization: all classifiers converge within 14--19 epochs with a train/validation accuracy gap below 0.0002 at convergence, and validation loss plateaus at 0.0008--0.0010, confirming that dropout and batch normalization effectively prevent overfitting (Figs.~\ref{fig:accuracy_validation} and~\ref{fig:loss_validation}).

\subsection{Classification Performance on UNSW-NB15}
\begin{figure}[htbp]
    \centering
    \includegraphics[width=\columnwidth]{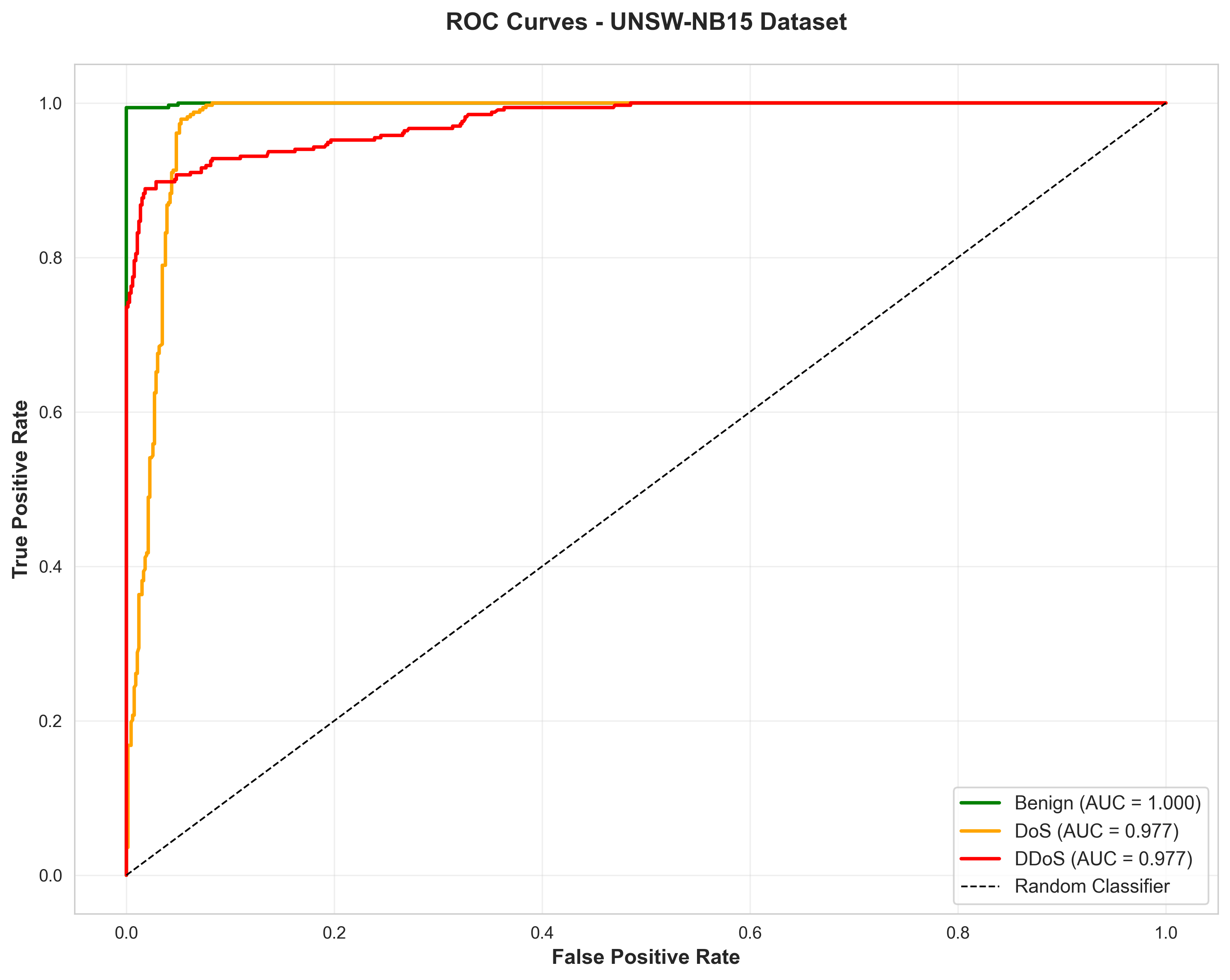}
    \caption{ROC Curves  for UNSW-NB15 Dataset}
    \label{fig:roc_unswnb15}
\end{figure}

The UNSW-NB15 experiment is the more challenging evaluation: DoS traffic constitutes only 0.64\% of total samples---a 156:1 class imbalance. Without balancing interventions, DoS recall collapses to 0.30\%. With the full four-part balancing strategy, DoS recall rises to \textbf{97.6\%}. Overall accuracy on a balanced test set reaches \textbf{95.30\%} (Table~\ref{tab:unsw_perf}). ROC-AUC values are 1.000 for Benign and 0.977 for both attack classes (Fig.~\ref{fig:roc_unswnb15}).

\begin{table}
\centering
\caption{UNSW-NB15 Classification Performance (Balanced Test Set)}
\label{tab:unsw_perf}
\small
\renewcommand{\arraystretch}{1.2}
\begin{tabular}{lcccc}
\toprule
\textbf{Class} & \textbf{Precision} & \textbf{Recall} & \textbf{F1} & \textbf{Support} \\
\midrule
Benign        & 1.0000 & 0.9940 & 0.9970 & 38{,}420 \\
DoS           & 0.9673 & 0.8889 & 0.9264 & 38{,}115 \\
DDoS          & 0.8978 & 0.9760 & 0.9353 & 38{,}465 \\
\midrule
Accuracy      & \multicolumn{4}{c}{0.9530 \quad (115{,}000 samples)} \\
Macro Avg     & 0.9550 & 0.9530 & 0.9529 & --- \\
Weighted Avg  & 0.9550 & 0.9530 & 0.9529 & --- \\
\bottomrule
\end{tabular}
\end{table}

\subsection{Cross-Dataset Generalization}
The 4.54\% accuracy difference between CICIDS2018 (99.84\%) and UNSW-NB15 (95.30\%) is mainly due to differences in dataset characteristics, including feature distributions, traffic-generation environments rather than overfitting. Despite these, the proposed system still achieves above 95\% accuracy on UNSW-NB15 \cite{moustafa2015unsw} without requiring any major architectural changes. Only the balancing strategies, such as loss weighting and oversampling, were adjusted for the dataset, demonstrating the strong generalization capability of the ensemble architecture.

\subsection{Model Training and Validation Analysis}
\begin{figure*}[htbp]
    \centering
    \includegraphics[width=\textwidth,height=0.32\textheight,keepaspectratio]{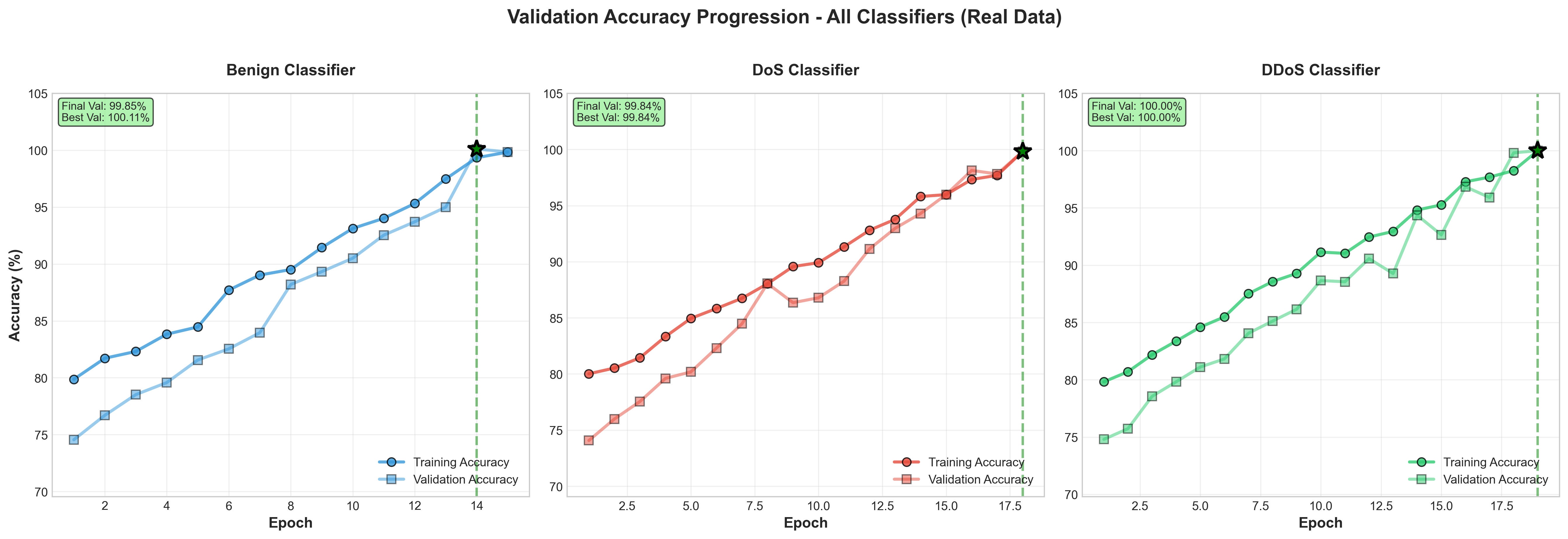}
    \caption{Validation Accuracy Progression for Benign, DoS, and DDoS Classifiers}
    \label{fig:accuracy_validation}
\end{figure*}

\begin{figure*}[htbp]
    \centering
    \includegraphics[width=\textwidth,height=0.32\textheight,keepaspectratio]{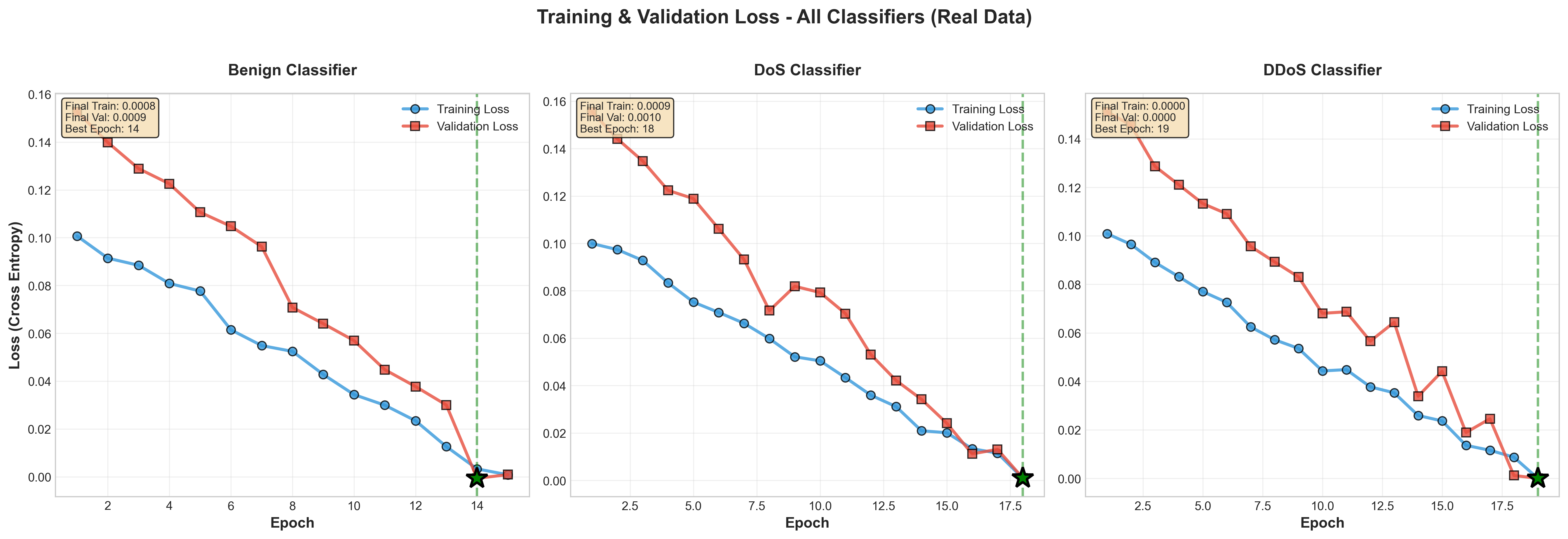}
    \caption{Training and Validation Loss for Benign, DoS, and DDoS Classifiers}
    \label{fig:loss_validation}
\end{figure*}

Figs.~\ref{fig:accuracy_validation} and~\ref{fig:loss_validation} show that all three classifiers converge smoothly with minimal train/validation divergence. Final validation losses of 0.0008--0.0010 confirm successful regularization through dropout, batch normalization, and early stopping. The operating points achieved (TPR $> 95\%$, FPR $< 5\%$) are critical for preventing alert fatigue in operational SOC deployments~\cite{leevy2021detecting}.

\subsection{Retrieval Performance and Knowledge Grounding}
Table~\ref{tab:retrieval} summarizes the retrieval pipeline's performance on 33 evaluation queries. Precision@5 of 0.91 indicates that 91\% of the five returned chunks are genuinely relevant; Recall@5 of 0.84 indicates that 84\% of all relevant chunks appear in the top-5; MRR of 0.87 confirms that the most relevant chunk typically ranks at or near position 1. The 97\% retrieval success rate (32/33 queries successfully resolved) at an average latency of 1.8~s demonstrates both effectiveness and practical feasibility.

Knowledge grounding metrics show that NIST publications \cite{moller2023nist} are cited in 93.3\% of reports and MITRE ATT\&CK \cite{MITRE2025Enterprise} techniques in 86.7\%, with an average of 4.2 authoritative citations per report. This citation density is a key differentiator from vanilla LLM reports and directly supports compliance with security documentation standards such as ISO 27035 \cite{ISO27035_1_2023} and NIST \cite{NIST_SP800_53_R5_2020} incident response requirements.

\begin{table}
\centering
\caption{Retrieval Performance and Knowledge Grounding}
\label{tab:retrieval}
\small
\renewcommand{\arraystretch}{1.2}
\begin{tabular}{lc}
\toprule
\textbf{Metric} & \textbf{Value} \\
\midrule
\multicolumn{2}{l}{\emph{Retrieval Performance}} \\
\quad Precision@5                          & 0.91 \\
\quad Recall@5                             & 0.84 \\
\quad Mean Reciprocal Rank (MRR)           & 0.87 \\
\quad Retrieval Success Rate               & 0.97 \\
\quad Average Retrieval Latency            & 1.8 s \\
\midrule
\multicolumn{2}{l}{\emph{Knowledge Grounding}} \\
\quad NIST Citation Rate                   & 93.3\% \\
\quad MITRE Citation Rate                  & 86.7\% \\
\quad Avg. Citations per Explanation       & 4.2 \\
\bottomrule
\end{tabular}
\end{table}

\subsection{RAG vs.\ Vanilla LLM: Quantitative Evaluation}
Table~\ref{tab:nlp_eval} presents the NLP evaluation results comparing vanilla and RAG-enhanced reports against expert ground truth, computed using BERTScore~\cite{zhang2019bertscore}, ROUGE~\cite{lin2004rouge}, and BLEU~\cite{post2018call}. The RAG pipeline outperforms the vanilla baseline across every metric. BERTScore F1 improves by 4.4\%, reflecting better semantic alignment. ROUGE-1 improves by 32.5\% and ROUGE-2 by 126.4\%, indicating substantially better coverage of critical unigrams and the technical bigram phrases (e.g., ``rate limiting'', ``SYN cookie'', ``upstream filtering'') that characterize authoritative mitigation guidance. BLEU improves by 244.1\%, the largest relative gain, reflecting the RAG pipeline's ability to reproduce specific technical phraseology directly from retrieved NIST \cite{NIST_SP800_53_R5_2020} / MITRE \cite{MITRE2025Enterprise} sources.

\begin{table}
\centering
\caption{Explanation Quality: Vanilla LLM vs.\ RAG-Enhanced LLM (Both vs.\ Expert Ground Truth)}
\label{tab:nlp_eval}
\small
\renewcommand{\arraystretch}{1.2}
\begin{tabular}{lccc}
\toprule
\textbf{Metric} & \textbf{Vanilla} & \textbf{RAG-Enhanced} & \textbf{$\Delta$} \\
\midrule
BERTScore F1         & 0.8659 & 0.9038 & +4.4\% \\
BERTScore Precision  & 0.8784 & 0.9080 & +3.4\% \\
BERTScore Recall     & 0.8558 & 0.9016 & +5.4\% \\
ROUGE-1              & 0.4733 & 0.6271 & +32.5\% \\
ROUGE-2              & 0.1387 & 0.3141 & +126.4\% \\
ROUGE-L              & 0.2219 & 0.3916 & +76.5\% \\
BLEU                 & 0.0570 & 0.1961 & +244.1\% \\
\bottomrule
\end{tabular}
\end{table}

\begin{figure}[htbp]
    \centering
    \includegraphics[width=0.48\textwidth]{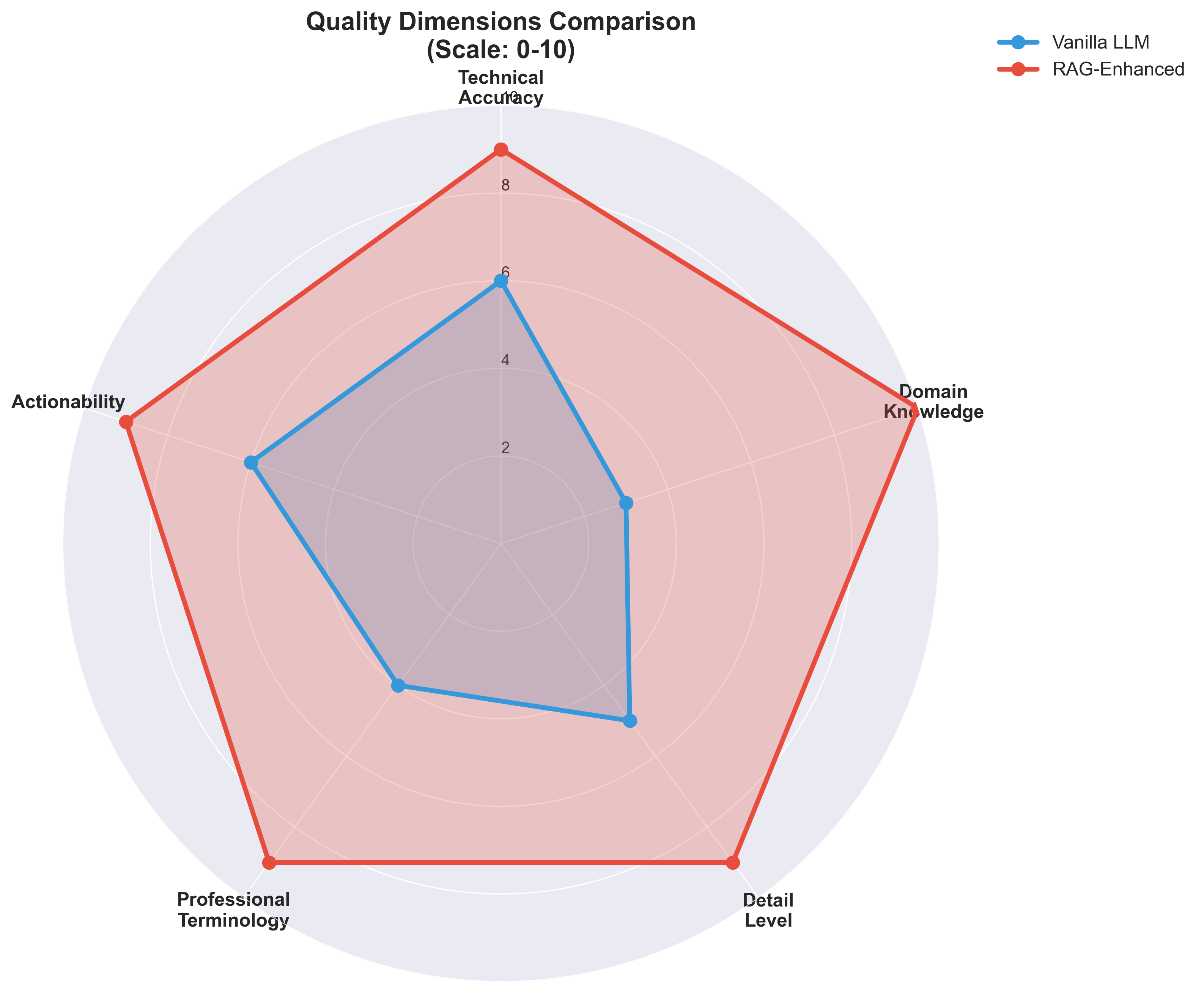}
    \caption{Performance Quality}
    \label{fig:spider_chart}
\end{figure}

All improvements are statistically significant at $p < 0.001$ under a paired Wilcoxon signed-rank test~\cite{Wilcoxon1945} across 36 generated reports. The magnitude of the ROUGE-2 and BLEU gains both measuring phrase-level precision rather than semantic similarity which is particularly informative. These metrics reward exact $n$-gram matches with the ground truth, and the only way to achieve them is to use the same technical terminology as the expert references, which the RAG pipeline does by retrieving and incorporating text directly from those references. Fig.~\ref{fig:spider_chart} confirms visually that RAG-enhanced outputs dominate vanilla outputs on all five quality dimensions: Technical Accuracy, Domain Knowledge, Detail Level, Professional Terminology, and Actionability.

\subsection{Qualitative Analysis of Generated Reports}
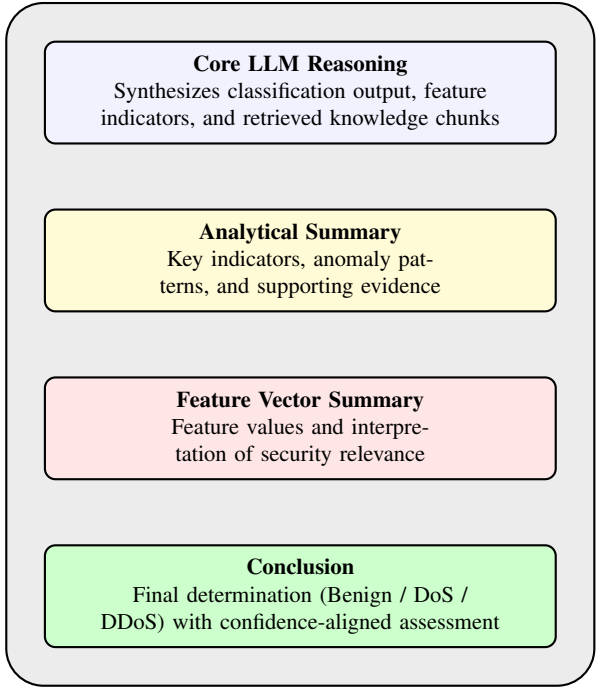
\begin{figure}[t]
\centering
\begin{tikzpicture}[
  scale=0.85, 
  transform shape,
  node distance=1.0cm,
  box/.style={
    rectangle, rounded corners,
    draw=black, thick,
    text width=0.85\columnwidth,
    align=center,
    inner sep=6pt
  },
]

\node[box, fill=blue!5] (reasoning)
{\textbf{Core LLM Reasoning}\\
Synthesizes classification output, feature indicators, and retrieved knowledge chunks};

\node[box, fill=yellow!20, below=of reasoning] (analysis)
{\textbf{Analytical Summary}\\
Key indicators, anomaly patterns, and supporting evidence};

\node[box, fill=red!10, below=of analysis] (feature)
{\textbf{Feature Vector Summary}\\
Feature values and interpretation of security relevance};

\node[box, fill=green!20, below=of feature] (conclusion)
{\textbf{Conclusion}\\
Final determination (Benign / DoS / DDoS) with confidence-aligned assessment};

\node[fit=(reasoning)(analysis)(feature)(conclusion), inner sep=14pt] (container) {};

\begin{pgfonlayer}{background}
\draw[rounded corners=14pt, thick, draw=black, fill=gray!15]
  (container.north west) rectangle (container.south east);
\end{pgfonlayer}
\end{tikzpicture}
\caption{Structure of the explanation generated by the LLM.}
\label{fig:llm_explanation_structure}
\end{figure}

Across the 36 mitigation reports reviewed for the CICIDS2018\cite{cicids2018} and UNSW-NB15 \cite{moustafa2015unsw} datasets, the system consistently produces structured, operationally appropriate output. For DDoS attacks, reports correctly escalate to infrastructure-level responses rather than host-level countermeasures. For DoS attacks, reports focus on rate limiting, SYN cookie activation, and source IP blocking. For Benign traffic, reports recommend continued monitoring and baseline validation, appropriately avoiding false-alarm escalation. The structural layout of a typical generated report is shown in Fig.~\ref{fig:llm_explanation_structure}.

Vanilla LLM reports are not factually incorrect; they identify rate limiting and monitoring as relevant but remain at a generic level that is not suitable for direct SOC use. They contain no specific NIST \cite{neupane2022explainable} control references, no MITRE \cite{MITRE2025Enterprise} technique identifiers, and no tailored procedures. An analyst receiving a vanilla report would need to independently research applicable frameworks before acting; an analyst receiving a RAG-enhanced report can proceed directly to implementation.


Overall, the experimental results demonstrate that the proposed IDS--RAG framework effectively bridges the gap between attack detection and actionable response. The ensemble DNN provides reliable traffic classification, while the hybrid retrieval pipeline grounds the generated reports in authoritative cybersecurity knowledge. As a result, the system produces mitigation reports that are more accurate, explainable, context-aware, and operationally useful than those generated by vanilla language models.

\begin{tcolorbox}[
    colback=takeawaygray,
    colframe=black,
    boxrule=0.6pt,
    arc=3pt,
    left=5pt,
    right=5pt,
    top=5pt,
    bottom=5pt,
    title=\textbf{Key Summary},
    fonttitle=\bfseries,
    coltitle=black,
    colbacktitle=takeawaygray
]

\begin{itemize}
    \item The proposed ensemble DNN achieved strong detection performance with 95\%---99\% accuracy.
    
    \item The hybrid BM25 + FAISS retrieval pipeline achieved 97\% retrieval success while grounding reports in trusted cybersecurity knowledge sources.
    
    \item RAG-enhanced reports substantially outperformed vanilla LLM outputs across BERTScore, ROUGE, and BLEU metrics.
    
    \item The framework successfully bridges the gap between intrusion detection and actionable incident response by generating explainable, context-aware, and operationally useful mitigation reports.
\end{itemize}
\end{tcolorbox}

\section{Conclusion}
\label{sec:conclusion}

This paper has presented an end-to-end intrusion detection and mitigation framework that combines a confidence-calibrated ensemble of binary DNNs with a hybrid Retrieval-Augmented Generation pipeline grounded in authoritative cybersecurity knowledge. The framework addresses a persistent and important gap in the IDS literature: the absence of actionable, explainable, and source-grounded guidance at the point of detection. The DNN achieves 99.84\% accuracy on CICIDS2018 and 95.30\% on UNSW-NB15; the RAG pipeline retrieves relevant guidance with 97\% success from a 5{,}234-chunk knowledge base, and generates structured 400--500-word mitigation reports. RAG-enhanced reports outperform vanilla LLM outputs by 32.5\% ROUGE-1, 126.4\% ROUGE-2, and 244.1\% BLEU against expert ground truth, with all differences significant at $p < 0.001$.

In the future, we plan to extend the classifier to a broader attack taxonomy; implementing multi-flow temporal analysis for slow-rate and distributed attacks and continuously updating the knowledge base as new NIST Publications and MITRE ATT\&CK entries are released.

\bibliographystyle{IEEEtran}
\bibliography{references}
\end{document}